\documentclass{article}

\usepackage{amsmath}
\usepackage{booktabs}
\usepackage{graphicx}
\usepackage[affil-it]{authblk}
\usepackage[caption=false]{subfig}

\usepackage{array}

\begin{document}
	
\title{Evaluation of network-guided random forest for disease gene discovery}

\author{Jianchang Hu
	\thanks{Email: \texttt{jianchang.hu@uni-luebeck.de}}}

\author{Silke Szymczak
	\thanks{Email: \texttt{silke.szymczak@uni-luebeck.de}; Corresponding author}}
\affil{Institute of Medical Biometry and Statistics,\\ University of Lübeck, Lübeck, Germany}

\maketitle

\begin{abstract}
\noindent \textbf{Motivation:} Gene network information is believed to be beneficial for disease module and pathway identification, but has not been explicitly utilized in the standard random forest (RF) algorithm for gene expression data analysis.\\
\textbf{Results:} We investigate the performance of a network-guided RF where the network information is summarized into a sampling probability of predictor variables which is further used in the construction of the RF. Our results suggest that network-guided RF does not provide better disease prediction than the standard RF. In terms of disease gene discovery, if disease genes form module(s), network-guided RF identifies them more accurately. In addition, when disease status is independent from genes in the given network, spurious gene selection results can occur when using network information, especially on hub genes. Our empirical analysis on two balanced microarray and RNA-Seq breast cancer datasets from The Cancer Genome Atlas (TCGA) for classification of progesterone receptor (PR) status also demonstrates that network-guided RF can identify genes from PGR-related pathways, which leads to a better connected module of identified genes.\\
\textbf{Availability:} https://github.com/imbs-hl/networkRF \\
\textbf{Contact:} silke.szymczak@uni-luebeck.de \\
\end{abstract}

\textbf{Keywords} Gene expression, Protein-protein interaction, RNA-Seq, Weighted random forest

\section{Introduction}

Gene expression analysis can quantify dynamic expression patterns under different biological conditions and thus help identify genes associated with complex diseases \cite{li2018review}. These biomarkers might improve patient risk prediction and can foster understanding of underlying molecular pathomechanisms. However, in the classic approach genes are often analyzed individually. Given the functional interdependencies between the molecular components, a complex disease such as cancer is rarely a consequence of an abnormality in a single gene \cite{goh2007, barabasi2011network}. Therefore, it would be beneficial to incorporate molecular interactions into analysis, where the interactions can be summarized in the form of molecular networks \cite{ideker2008network}. Examples include protein-protein interaction (PPI) networks and gene regulatory networks where each node of these networks represents a gene and the edges between nodes reflect their interactions (see \cite{barabasi2011network} for a review on different types of biological networks). In this regard, gene network information is believed to be beneficial for disease module and pathway identification where a disease module or pathway is expected to be a local cluster of highly connected genes within the network.

To identify disease associated genes, many statistical and machine learning approaches have been developed \cite{li2018review}. One popular machine learning method is the random forest (RF) algorithm \cite{breiman_random_2001}. It is a nonparametric approach that can accommodate different types of phenotypes including categorical or quantitative phenotypes and survival times \cite{rfsurvival}. Moreover, it can work with predictors of various scales or distributions and is suited for applications in high-dimensional settings such as transcriptomics data where the number of predictors can be larger than the number of observations \cite{chen_random_2012, cutler_random_2007}. With the so-called variable importance measures, the algorithm can also highlight the relevance of each predictor to the prediction of phenotype \cite{breiman_random_2001}. This is especially handy for disease gene discovery where genes associated with disease phenotypes can be identified.

However, as with many other methods for transcriptome analysis, the standard RF algorithm only uses gene expression data for constructing the prediction model and identifying disease genes. The gene network information has not been explicitly utilized. Therefore, in this paper, we investigate a network-guided RF where the network information is summarized into a sampling probability of predictors which is further used in the construction of RF. This sampling probability can be considered as prior knowledge on the importance of each predictor on model construction. In the standard RF, this sampling probability of predictors is a uniform probability to reflect that we do not impose any prior knowledge on the importance of predictors. However, it is possible that we have certain prior knowledge or belief on the importance of predictors from external sources such as gene network. For instance, hub genes in the gene network tend to play more important roles in disease progression \cite{barabasi2011network}. Therefore, we may want to increase the usage frequency of these genes in the construction of RF. This can be achieved by modifying the sampling probability of predictors. 

The strategy to modify the sampling probability of predictor variables in RF to prioritize certain predictors is not unknown in the bioinformatics literature. The Enriched RF \cite{amaratunga2008enriched} uses the false discory rate (FDR) adjusted p-values from marginal association tests of each predictor to construct such sampling probabilities. The variable importance-weighted RF \cite{liu2017} first constructs a standard RF and obtains the variable importance measures of all predictors. The estimated variable importance measures are then normalized to be the sampling probabilities of predictors in the second-round RF construction.

The network-guided RF also adopts this strategy. Different from the Enriched RF and variable importance-weighted RF, it always considers network information as part of the prior knowledge to prioritize predictors. One example of network-guided RF is given by Wang and Liu (2018) \cite{wang2018SciRepo} where they construct a variant of the random survival forest to build a better prediction model by selecting genes that show a great ability to predict the survival endpoint. In their RF construction, the sampling probability of predictors is based on p-values of marginal association tests and gene network information. 
Furthermore, the aforementioned approaches are constructed for a better predictive model instead of more accurate variable selection. Especially for the RF in  Wang and Liu (2018), it is only applied on experimental data, no simulation studies have been designed to investigate its prediction and variable selection performance. Therefore, it is necessary to evaluate the network-guided RF from the disease gene identification perspective. Specifically, we would like to understand \emph{when the network information is beneficial for identifying disease genes and modules}.

We conduct simulation studies to evaluate the performance of network-guided RF in terms of disease gene and module identification as well as prediction accuracy. We analyze the effect of incorporating information from marginal association tests or networks separately and together. A module-based synthetic gene network is first simulated and RNA-Seq data are then simulated based on the synthetic network. We further apply the method utilizing a protein-protein interaction network to identify genes relevant for classifying progesterone receptor (PR) status in breast cancer patients. We consider PR status of breast cancer because it has been extensively studied in the genomic literature and several molecular pathways are well known \cite{lange2008progesterone}. We use two independent breast cancer datasets from The Cancer Genome Atlas (TCGA) that were generated using microarray and RNA-Seq, respectively. We consider the concordance between these results as a validation for the discovery.

\section{Materials and methods}

\subsection{Network-guided RF}

In this paper, we work with binary phenotypes such as disease status, but the network-guided RF can be adopted for other types of responses as well.

For a binary classification problem, the standard RF consists of an ensemble of binary classification and regression trees (CARTs) \cite{breiman1984classification} where each tree is built from a bootstrapped version of the training data. Each tree is grown via the principle of recursive partition where starting from the root node, the same node splitting procedure is applied recursively until certain stopping rules are met. The node splitting procedure consists of selecting a splitting variable and determining the splitting rule. To select the splitting variable, a pre-determined number of candidate splitting variables are randomly selected from all predictors. Each of these randomly sampled predictors is then investigated to search for the best splitting variable and to determine the splitting rule. The guiding principle for node splitting is to minimize the impurity of response values in each node, which is often measured by the Gini index for classification problem.

The network-guided RF is a variant of the standard RF to incorporate gene network information into RF construction for gene expression analysis. It achieves this by modifying the sampling probability of predictor variables during the node splitting procedure. In the standard RF, the node splitting procedure starts with randomly sampling a subset of predictor variables to be investigated for node splitting at a given node. The sampling probability is uniform, meaning that all predictors are equally likely to be selected as candidate splitting variables. However, with a given gene network, we have information on their topological importance. For instance, hub genes, which are connected to many other genes in the network, may have a higher importance. Hence, we can modify this sampling probability to reflect such network-based importance information and to prioritize the use of these genes in the construction of RF in the hope that it can help us identify disease modules and genes more efficiently. The network-guided RF takes this approach to incorporate the network information into the construction of RF.

In the network-guided RF, the creation of this sampling probability with a given gene network is based on the directed random walk (DRW) algorithm \cite{kohler2008drw}. The core idea of the algorithm is to simulate a random walker on the given network which starts at a source node and, at each step, with probability of $1-r$ it moves from the current node to a randomly selected neighboring node, or with a restart probability of $r$ it goes back to the source node. After a number of steps, the probability distribution of the random walker being at each node in the network will reach an equilibrium and this stabilized distribution thus reflects the topological importance of genes with respect to the initial source node in the given gene network.

Mathematically, let $A$ be the row-normalized adjacency matrix of the given gene network, where an adjacency matrix of a given network is a square matrix with the size being the same as the number of nodes in the network and with elements
\[ A_{ij} = \mathcal{I}(\text{there exists an edge between node } i \text{ and } j),\ \forall i \neq j, \]
where $\mathcal{I}(\cdot)$ is the indicator function. The DRW iterations are given by
\begin{equation}\label{eqn:DRW_iterates}
	\pi_{t+1} = (1-r) A^\top \pi_t + r \pi_0,
\end{equation}
where $\pi_t$ is a vector whose $i$th element is the probability of the random walker being at node $i$ at iteration $t$, and $\pi_0$ is the initial distribution over the network, and $r$ is the pre-determined restart probability. After a number of iterations, $\pi_t$ becomes stable and can be considered as converged to an equilibrium distribution $\pi^\ast$.

If $r=0$, then there is no restart and this algorithm reduces to the standard random walk over the network and the resulting equilibrium probability of each node is independent of the initial distribution $\pi_0$, and is often proportional to the degree of the node, i.e., the more connections the node has to other nodes, the higher chance for the random walker to stay at the node. When non-zero restart probability $r$ is adopted, the resulting equilibrium deviates from the aforementioned equilibrium distribution of standard random walk and is influenced by the initial distribution $\pi_0$. In particular, as in \cite{wang2018SciRepo}, if $\pi_0$ is constructed by assigning $-\log(p_i)$ as its $i$th element where $p_i$ is the $p$-value of marginal association test of gene $i$ and normalizing it to a unit vector, then in the equilibrium distribution, genes that have large degrees in the network, have significant $p$-values and are closer to the previous two categories of genes will have higher probabilities.

In short, the construction of network-guided RF can be summarized as follows.
\begin{enumerate}
	\item[(1)] For a given gene network and any other external information, use the external information to construct an initial distribution $\pi_0$ over all genes and run the DRW algorithm on the given network with iterations according to equation \eqref{eqn:DRW_iterates} to obtain the (approximate) equilibrium distribution $\pi^\ast$. 
	\item[(2)] Draw \texttt{ntree} bootstrap samples, i.e. draw with replacement, from the gene expression dataset used for training.
	\item[(3)] For each bootstrap sample, grow a CART-based decision tree. During the tree construction process, at each node, \texttt{mtry} genes are randomly selected according to distribution $\pi^\ast$ as candidate splitting variables. The default value of \texttt{mtry} is $\sqrt{p}$ where $p$ is the total number of genes in the training set.
	\item[(4)] Grow each tree to full size until pre-determined stopping rules are met (e.g., minimum node size or complete purity of the node).
	\item[(5)] The nodes in the final layer of a tree are used for prediction of new observations. To make prediction with the RF, an observation goes through every decision tree in the forest and the final prediction for the observation is made based on aggregating results from all decision trees in the forest.
\end{enumerate}

\subsection{Selection of important genes}

The so-called variable importance measure can be obtained from the network-guided RF. Specifically, we consider the permutation-based variable importance measure \cite{breiman_random_2001} to evaluate the importance of genes. Based on the variable importance, the most important genes are selected using an approach similar to the one in \cite{wang2018SciRepo}. Starting with all genes in the dataset, at each step, the network-guided RF is constructed and the 10\% lowest ranking genes are discarded. The remaining 90\% of genes are used to construct the network-guided RF at the next step until a pre-determined minimum number of genes is retained.
This selection procedure shares similar idea with recursive feature elimination (RFE) \cite{diaz2006rfe} and the only difference is that the number of genes to be retained is pre-determined in the current approach while this number is based on the prediction performance of the model in RFE.
These selected genes are then considered as most important and relevant genes for the disease phenotype.

\subsection{Simulation study}

Here we describe our simulation study to evaluate the network-guided RFs. The description is structured according to the ADEMP scheme \cite{morris2019ademp}.

\subsubsection{Aim}

The simulation study is conducted to systematically evaluate the network-guided RFs on disease classification and disease gene identification accuracy under various scenarios.

\subsubsection{Data generation}

We generate synthetic gene expression data along with its underlying network structure using the R package \texttt{SeqNet} Version 1.1.3 \cite{grimes2021seqnet}. The network with a given number of genes is first randomly generated and then kept fixed for all scenarios. The network follows a module-based construction where several small modules are first generated. Then these modules are connected in a way that the degrees of all genes follow a power law which is observed in many biological networks \cite{milo2002network}. With a synthetic network, the simulator then generates data from a Gaussian graphical model and convert those values into RNA-seq gene expression data. The marginal distribution of expression for each gene is calibrated from a reference TCGA breast cancer RNA-seq dataset. More details of the data generation procedure can be found in \cite{grimes2021seqnet}.

The binary disease status is generated from a logistic regression model. In particular, the phenotype follows a Bernoulli distribution and the log-odds is modeled by the following equation.
\begin{equation}\label{eqn:sim_model}
	\text{log-odds} = \beta_0 + \sum_{i\in D} \beta_i X_i,
\end{equation}
where set $D$ denotes the indices of disease genes, $X_i$ is the standardized gene expression data of disease gene $i$ with mean 0 and standard deviation 1, and $\beta_0$ is the intercept and $\beta_i$'s are regression coefficients to reflect the effect sizes of disease genes. In our simulation, we set $\beta_0 = 0$.

We consider various scenarios in the simulation and they are summarized in Table \ref{tab:sim_setup}.

\begin{table}[h]
	\centering
	\caption{Summary of different scenarios in simulation studies. Scenarios are ordered by the number of disease modules. Information on how the effect sizes of disease genes distribute within a disease module is also given.}
	\begin{tabular}{lp{0.25\linewidth}p{0.45\linewidth}}
		\toprule
		Scenario & Number of disease modules & Distribution of effect sizes of genes within disease modules \\
		\midrule
		Null & 0 & None \\             				
		RanEqu & 0 & Uniform \\             				
		ModEqu & 1 & Uniform \\             				
		ModTopo & 1 & Connectivity-based \\             				
		TwoModEqu & 2 & Uniform \\             				
		TwoModTopo & 2 & Connectivity-based\\
		\bottomrule
	\end{tabular}
	\label{tab:sim_setup}
\end{table}

\begin{figure}[h]
	\includegraphics[width=0.48\columnwidth]{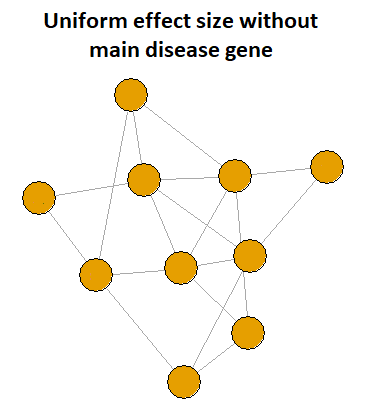}
	\includegraphics[width=0.5\columnwidth]{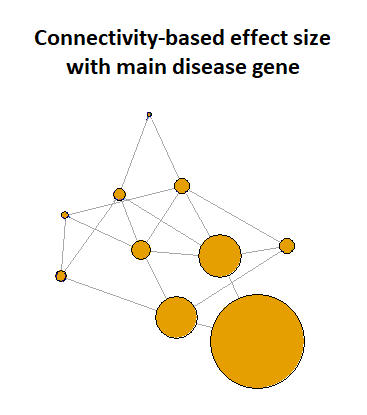}
	\caption{Illustration of disease modules with and without main disease gene. The size of the bubble reflects the effect size of the gene. When there is no main disease gene (as shown on the left), all disease genes have the same effect size. When there is a main disease gene (as shown on the right), the effect size of each disease gene is proportional to its closeness to the main disease gene within the module; the closer to the main disease gene, the larger the effect size.}
	\label{fig:disease_module}
\end{figure}

We have the Null case where no gene in the network is relevant for the disease status. This scenario gives us the opportunity to investigate the performance of network-guided RFs in terms of false selection of genes. In scenario RanEqu, disease genes are randomly distributed in the network and they have the same effect size, i.e., $\beta_i = \beta, \forall i\in D$. In contrast, scenarios ModEqu and ModTopo consider a disease module where all disease genes come from one topological module in the network. The difference between these two scenarios is the way to assign effect sizes to these disease genes. In ModEqu all disease genes have the same effect size while in ModTopo, there is a main disease gene randomly selected within the module and the effect sizes of all other disease genes are then based upon the topological closeness to the main disease gene. Figure \ref{fig:disease_module} gives an illustration of the disease modules and the assignment of effect sizes in these two scenarios. Scenarios TwoModEqu and TwoModTopo further extend these disease module-based scenarios to allow two different non-overlapping disease modules.

The number of disease genes in RanEqu and each disease module is approximately the first quartile of the sizes of all modules in the network generated by \texttt{SeqNet} which in our simulations is around 12-20 genes while the total number of genes is $p\in \{1000, 3000\}$. We consider three different average effect size $\beta=\{0.5, 1, 2\}$. For scenarios with equal effect sizes, $\beta_i = \beta$ for all $i\in D$. For disease modules with a main disease gene, the DRW algorithm with initial distribution concentrated only on the main disease gene (i.e., the distribution has probability one on the main disease gene and zero elsewhere) is first run to obtain an equilibrium distribution over the module. Then the effect sizes are calculated as $\beta_i = \pi_i^\ast * \beta * |M|$ where $\pi_i^\ast$ is the equilibrium probability of gene $i$ in the disease module from the DRW algorithm and $|M|$ gives the number of genes in the disease module $M$. In this way, the coefficients in equation \eqref{eqn:sim_model} always add up to $\beta * |D|$ and thus, $\beta$ is the average effect size.

For each scenario and combination of $(p, \beta)$, we simulate 100 independent replications with 2000 independent samples. Because we set $\beta_0=0$ in the logistic model in equation \eqref{eqn:sim_model}, the case-to-control ratio is on average 1. In each replication, the whole dataset is further equally split into training and testing sets with each having 1000 samples.

\subsubsection{Estimand}

On the training set, each method estimates the importance of all genes and thus gives a ranking of genes. We calculate the permutation importance measure in all RF constructions. Based on the ranking, top $|D|$ most important genes are identified and they are further used to build a RF model for prediction. In addition, we also record the number of times each gene is selected as one of the top genes to assess the disease gene identification performance.

\subsubsection{Methods to be evaluated}

We compare the following methods.
\begin{enumerate}
	\item Oracle: standard RF with only disease genes. This serves as the benchmark for all other methods.
	\item Standard RF: standard RF with all genes in the dataset.
	\item Marginal-P: sampling probability of genes in RF construction is based on the $p$-values of marginal two-sample $t$-tests per gene, i.e., the initial distribution $\pi_0$ used in \cite{wang2018SciRepo} as we discussed in the previous section.
	\item Network-guided RFs:
	\begin{itemize}
		\item Topology: sampling probability of genes is based on the network structure only to reflect the gene network topology, i.e., $r=0$ in the DRW algorithm.
		\item Sampling probability of genes is based on marginal association information and network topology. If the $p$-values of marginal two-sample $t$-tests are used, it is denoted as Network-P. If the FDR-adjusted $p$-values of marginal two-sample $t$-tests are used, it is denoted as Network-Q. We fix the restart probability in the DRW algorithm to be $r=0.3$ as in \cite{wang2018SciRepo}.
	\end{itemize}	
\end{enumerate}

All RF implementations are based on the R package \texttt{ranger} Version 0.14.1 \cite{marvin2017ranger}. In the simulation, we set the number of trees to be 1000 and \texttt{mtry} to be the default value of \texttt{ranger} (i.e., at each node, we randomly sample $\sqrt{p}$ genes as candidate splitting variables where $p$ is the total number of genes in the training dataset). An R package \texttt{networkRF} has been prepared to include codes for all methods and simulation scenarios considered in our study. More information on the R package can be found in the section Data availability.

\subsubsection{Performance measure}

The performance of all methods is evaluated in the following three aspects:
\begin{itemize}
	\item Disease prediction: In each replication, each method builds a RF predictive model. This model is applied to the testing set to obtain the misclassification rate. The average misclassification rate is calculated based on all 100 replications.
	
	\item False selection: In the Null case, the frequency of each gene being selected as important gene over all 100 replications is recorded to assess the false selection performance of all methods.
	
	\item Sensitivity to select disease genes: In scenarios where there are disease genes, the average proportion of disease genes being selected by each method is reported as the performance measure for disease gene selection.
\end{itemize}

\subsection{Experimental datasets}

For the experimental data, we use two preprocessed and curated TCGA breast cancer gene expression datasets generated with two different technologies (microarray and RNA-sequencing) for prediction of progesterone receptor (PR) status that are obtained from the Bioconductor package \texttt{curatedTCGAData} \cite{ramos2020curatedTCGA}. Only data from primary tumors are used. We remove replicates and only keep the first observation and remove genes with more than 50\% zero counts. We retain only white female patients and create two non-overlapping and balanced datasets. We have 283 and 284 patients for microarray and RNA-Seq datasets, respectively. Both datasets include 193 PR positive patients. Gene expression values are standardized to have mean 0 and standard deviation 1 per gene in each data set separately (same as in \cite{degenhardt2019evaluation}).

Besides the gene expression data, we also use PPI network information from the STRING database \cite{szklarczyk2023string} via the R package \texttt{STRINGdb} \cite{franceschini2015stringdb}. We use STRING version 11.5 for human species and adopt default quality control on the interactions by the package. We further remove proteins with duplicated STRING ID and duplicated mapped gene ID. In total, we retain 14167 common genes in both datasets along with the corresponding PPI network consisting of 1,290,904 interactions. Please refer to the Data availability section for the access to the experimental datasets.

In the experimental data analysis, we do not include method Oracle because the disease genes are unknown. We also do not include method Network-Q for comparison because it considers similar information sources as Network-P. Therefore, we have in total four methods for comparison, namely standard RF, Marginal-P, Topology and Network-P where the first two methods do not use the PPI information while the last two incorporate network information into RF construction.

On both microarray and RNA-Seq datasets, each method constructs a RF with 10,000 trees (the same number of trees used in \cite{degenhardt2019evaluation} and \cite{seifert2019surrogate} on similar datasets) and retains top 200 important genes and uses them to build a RF predictive model. Given the results in \cite{seifert2019surrogate}, we think this number is a reasonable choice for our datasets. The models built with microarray data are then tested on RNA-Seq data and vice versa to give the prediction accuracy.

\section{Results}

\subsection{Simulation results}

\subsubsection{Prediction accuracy}

We first look at the prediction accuracy of all methods under the different simulated scenarios. 

\begin{figure}[!h]
	\centering
	\includegraphics[width=\columnwidth]{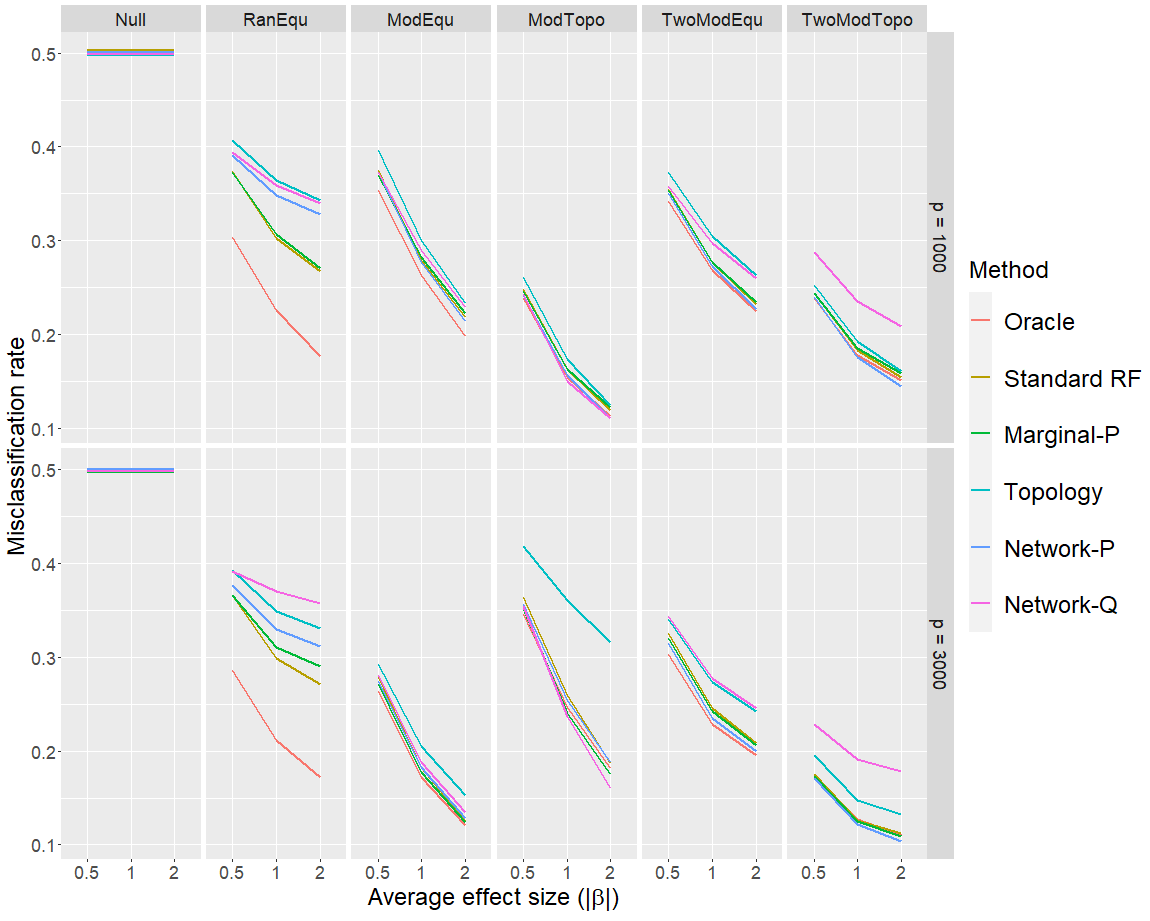}
	\caption{Prediction performance of all methods in all simulation scenarios. The performance is measured by average misclassification rate calculated on the testing set over 100 repetitions. In each scenario, three average effect sizes are considered to represent the cases for weak, median and strong signals. The upper panel gives the results for $p=1000$ total number of genes which is the same as the number of training samples, and the lower panel gives the results for $p=3000$ total number of genes to represent the case of high-dimensional setting.}
	\label{fig:sim_res_misclassification}
\end{figure}

As shown in Figure \ref{fig:sim_res_misclassification}, in the Null case, all methods are essentially performing random guess, so the misclassification rates are 0.5 as expected. In all other scenarios, Oracle has the best accuracy because it uses only the disease genes to build the predictive model, thus it serves as the benchmark for prediction accuracy.

When there exist disease genes, all methods have higher prediction accuracy as average effect size increases as expected. In addition, the prediction accuracy of all methods is not significantly affected by the different total number of genes, showing the advantage of RF-based methods in high dimensional settings.

When disease genes are not in any module as in the RanEqu scenario, network-guided RFs are clearly outperformed by methods without using network information. In fact, standard RF has the best accuracy in this scenario.

When disease genes are from disease module(s), the network-guided RFs show competetive prediction accuracies. This is especially true for Network-P where network information combined with $p$-values of marginal tests are used. The misclassification error of Network-P is among the lowest in most module-based scenarios, but the improvement over standard RF and Marginal-P is not substantial. The results for Network-Q, which also considers both network and marginal association information, are unstable, the method has comparable accuracy when there is only one disease module, but performs undesirably when there are two disease modules. The Topology method where only network topology information is used performs consistently undesirable with the largest error in most scenarios.

Overall, from the prediction perspective, network information does not add much value. This is not completely surprising because genes within a module are usually correlated with each other.
Using all disease genes in module-based scenarios to construct predictive model may not be necessary. This is also manifested by the comparison with Oracle. Combined with the variable selection results in the next subsection, we can see that several methods can achieve near-optimal prediction accuracy in these scenarios without using all disease genes.

\subsubsection{Disease gene identification}

Now we look at disease gene identification and start with the Null case because this provides a clue on the number of falsly selected genes.

\begin{figure}[!h]
	\centering
	\includegraphics[width=\columnwidth]{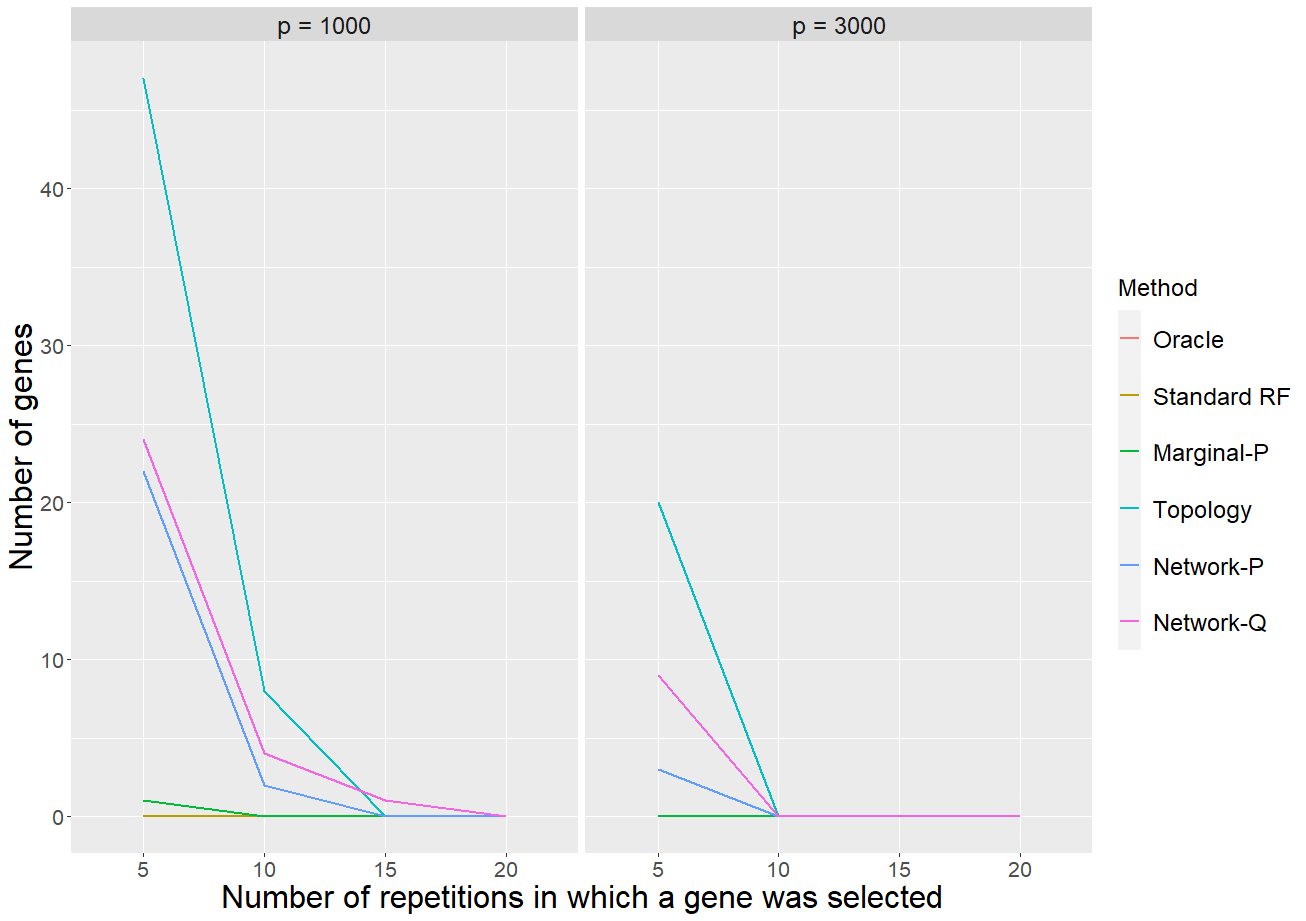}
	\caption{Number of genes being consistently selected as important genes by each method in the null case. We demonstrate this consistency by counting the number of repetitions within 100 repetitions that a given gene is selected as important genes. The plot shows counts of false selection at several consistency threshold level.}
	\label{fig:sim_res_false_pos}
\end{figure}

In Figure \ref{fig:sim_res_false_pos}, we record the number of genes that have been consistently selected as important genes by each method in 100 repetitions. In Null case no gene should be relevant to the phenotype, so we expect that no gene should be consistently ranked at the top. However, as we can clearly see in Figure \ref{fig:sim_res_false_pos}, network-guided RFs have a tendency to consistently rank a subset of genes as the top important genes. A closer look at these genes reveals that this subset of genes concentrates on hub genes which are highly connected genes in the network. This is largely because the sampling probability prioritizes the use of hub genes in the RF construction. This increased usage frequency artificially creates the spurious importance when no gene is relevant to the phenotype. However, from our simulation, the good news is that when the total number of genes in the dataset increases, this phenomenon of consistent false selection becomes less severe.

Next we look at the sensitivity of all methods to select disease genes. Because the Oracle by design only uses disease genes, its sensitivity is always 1.

\begin{figure}[!h]
	\centering
	\includegraphics[width=\columnwidth]{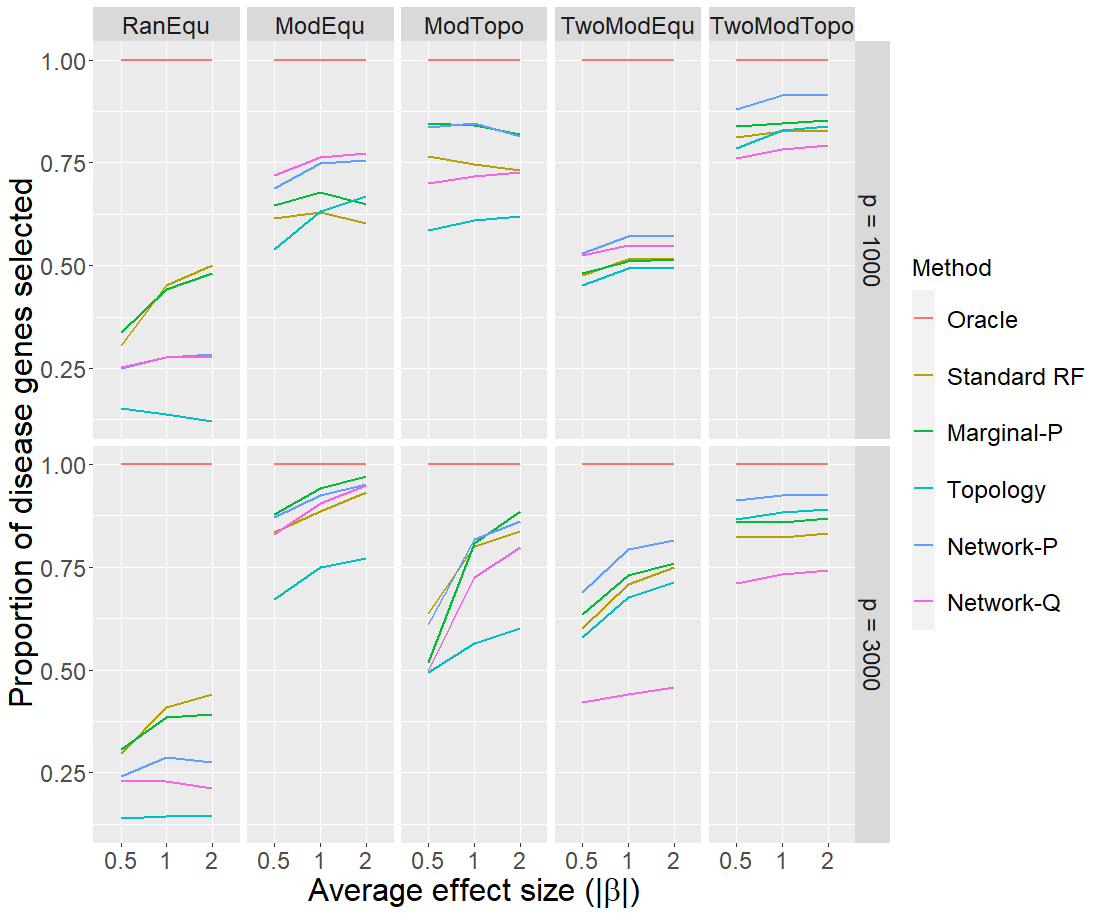}
	\caption{Sensitivity to select disease genes of all methods in all simulation scenarios. The performance is measured by average proportion of disease genes selected as important genes by each method over 100 repetitions. In each scenario, three average effect sizes are considered to represent the cases for weak, median and strong signals. The upper panel gives the results for $p=1000$ total number of genes which is the same as the number of training samples, and the lower panel gives the results for $p=3000$ total number of genes to represent the case of high-dimensional setting.}
	\label{fig:sim_res_sensitivity}
\end{figure}

Figure \ref{fig:sim_res_sensitivity} shows that when average effect size increases, all methods improves on the sensitivity in general. When the disease genes are not in modules as in the scenario RanEqu, methods without using network information outperform network-guided RFs by a great margin. This is because network-guided RFs assign weights based on the network topology which in this case is irrelevant. Furthermore, in this scenario, Topology is always the worst as it puts too much weights on hub genes which are not necessarily disease genes.

When disease genes form module(s), the selection accuracy improves a lot, especially for Network-P. In most scenarios with disease modules, Network-P is one of the top performers. Network-Q again shows unstable performance. This could due to the adjustment it makes to the $p$-values which leads the method to put too much weight on only a small number of genes. The consequence is that, for instance, when there are two disease modules, the method has a higher chance to miss one disease module or can only capture the main disease genes within a disease module.

It is also interesting to observe that even though the difference between standard RF and Marginal-P is not huge, incorporating marginal association information does help to achieve a slightly better sensitivity in most cases.

\subsection{Breast cancer datasets results}

Now we look at the results for experimental data analysis. We start with the prediction performance of all methods which are summarized in Table \ref{tab:pred_accuary}. The numbers reported are the misclassification errors obtained on the corresponding data with the model trained with the dataset from the other technology. For instance, the standard RF trained by RNA-Seq data achieves 13.03\% misclassification rate when it is applied to microarray data for testing.

\begin{table}[h]
	\centering
	\caption{Prediction performance of all methods on the experimental datasets. The reported numbers are the misclassification errors on the corresponding data in the header with the model trained on the dataset with the other technology.}
	\begin{tabular}{lcc}
		\toprule
		Method & Microarray & RNA-Seq \\
		\midrule            				
		Standard RF & 13.03\% & 14.13\% \\             				
		Marginal-P & 12.68\% & 13.43\% \\             				
		Topology & 11.97\% & 13.07\% \\             				       				
		Network-P & 12.32\% & 12.37\% \\
		\bottomrule
	\end{tabular}
	\label{tab:pred_accuary}
\end{table}

Recall that the proportion of positive PR status is 0.682 and 0.680 in microarray and RNA-Seq dataset respectively, meaning that all methods are substantially better than random guessing. We also find that models trained with RNA-Seq data enjoy better prediction performance on microarray data than the other way around. While all methods have similar prediction accuracy, network-guided RFs do provide slightly better misclassification rates on both datasets.

\begin{figure}[!h]
	\centering
	\includegraphics[width=\columnwidth]{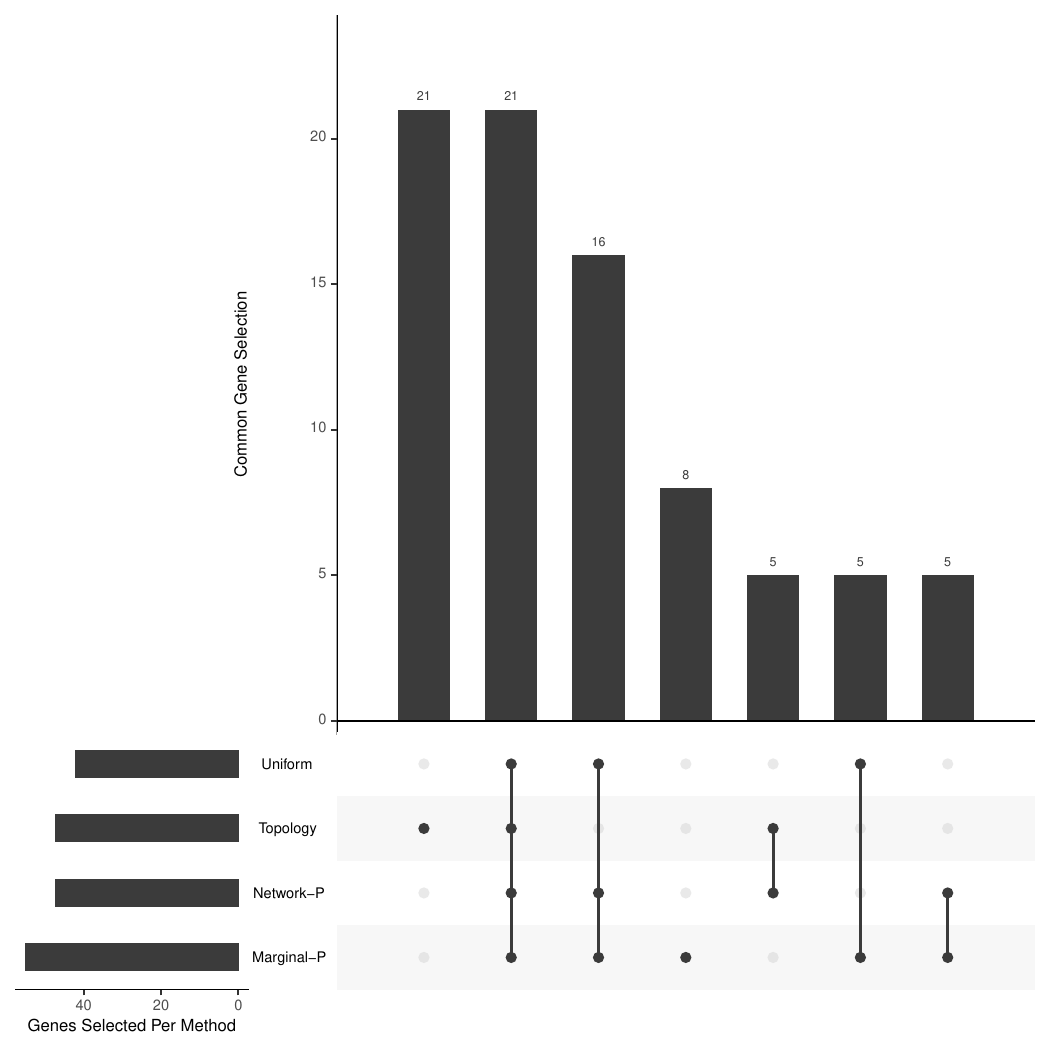}
	\caption{Common top genes selected by each method on both TCGA breast cancer microarray and RNA-Seq datasets.}
	\label{fig:exp_venn_common}
\end{figure}

Next we look at the identified genes, and the results here are only descriptive. We first look at the common genes that are ranked at the top by all four methods on both datasets. The UpSet diagram in Figure \ref{fig:exp_venn_common} shows the common genes that are ranked at the top by all four methods on both TCGA breast cancer microarray and RNA-Seq datasets in the experimental data analysis.

We can see that in total there are 21 genes that are considered by all methods on both datasets as important genes. 

\begin{figure}[!h]
	\centering
	\vspace{-0.5in}
	\includegraphics[width=\columnwidth]{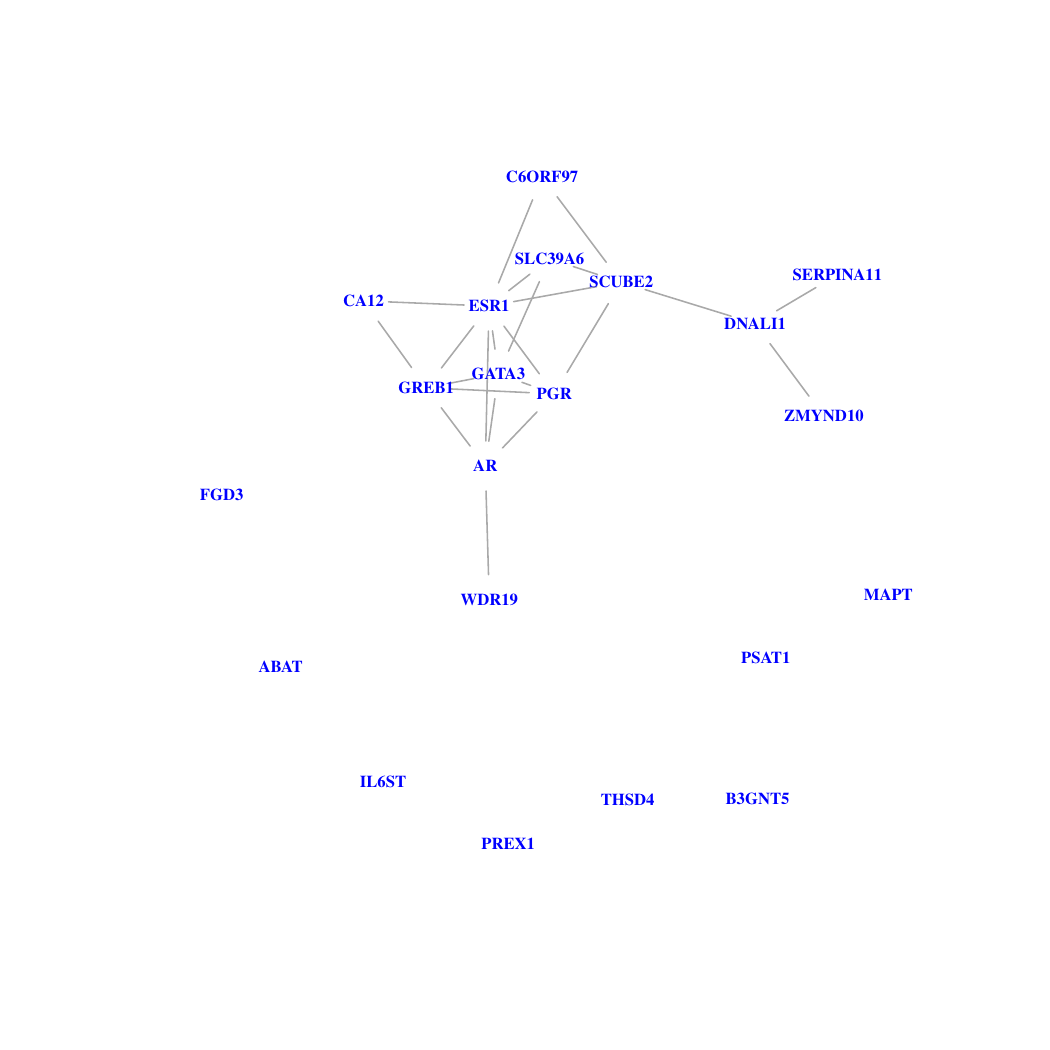}
	\vspace{-0.5in}
	\caption{Induced network of 21 common top genes selected by all 4 methods on both microarray and RNA-Seq datasets.}
	\label{fig:exp_network_common_all}
\end{figure}

The induced network of these 21 genes is shown in Figure \ref{fig:exp_network_common_all}. The network consists of a connected module of 13 genes and 8 isolated genes. As shown in the figure, the module contains the core PR status related progesterone receptor (\emph{PGR}) gene. In addition, gene \emph{PGR} is known to be related to estrogen receptor-mediated (ESR-mediated) signaling and Gene expression (Transcription) pathways. In this induced network, we can find genes Estrogen Receptor 1 (\emph{ESR1}),  GATA Binding Protein 3 (\emph{GATA3}) and Growth Regulating Estrogen Receptor Binding 1 (\emph{GREB1}) from ESR-mediated signaling pathway and Androgen Receptor (\emph{AR}) gene from Gene expression (Transcription) pathway. Gene \emph{SCUBE2}, short for Signal Peptide, CUB Domain And EGF Like Domain Containing 2, has also been shown to play an important role in breast cancer progression, especially for the triple-negative breast cancer subtype \cite{kumar2022scube}.

\begin{figure}[htp]
	\vspace{-1.5in}
	\subfloat[]{%
		\includegraphics[width=0.85\columnwidth]{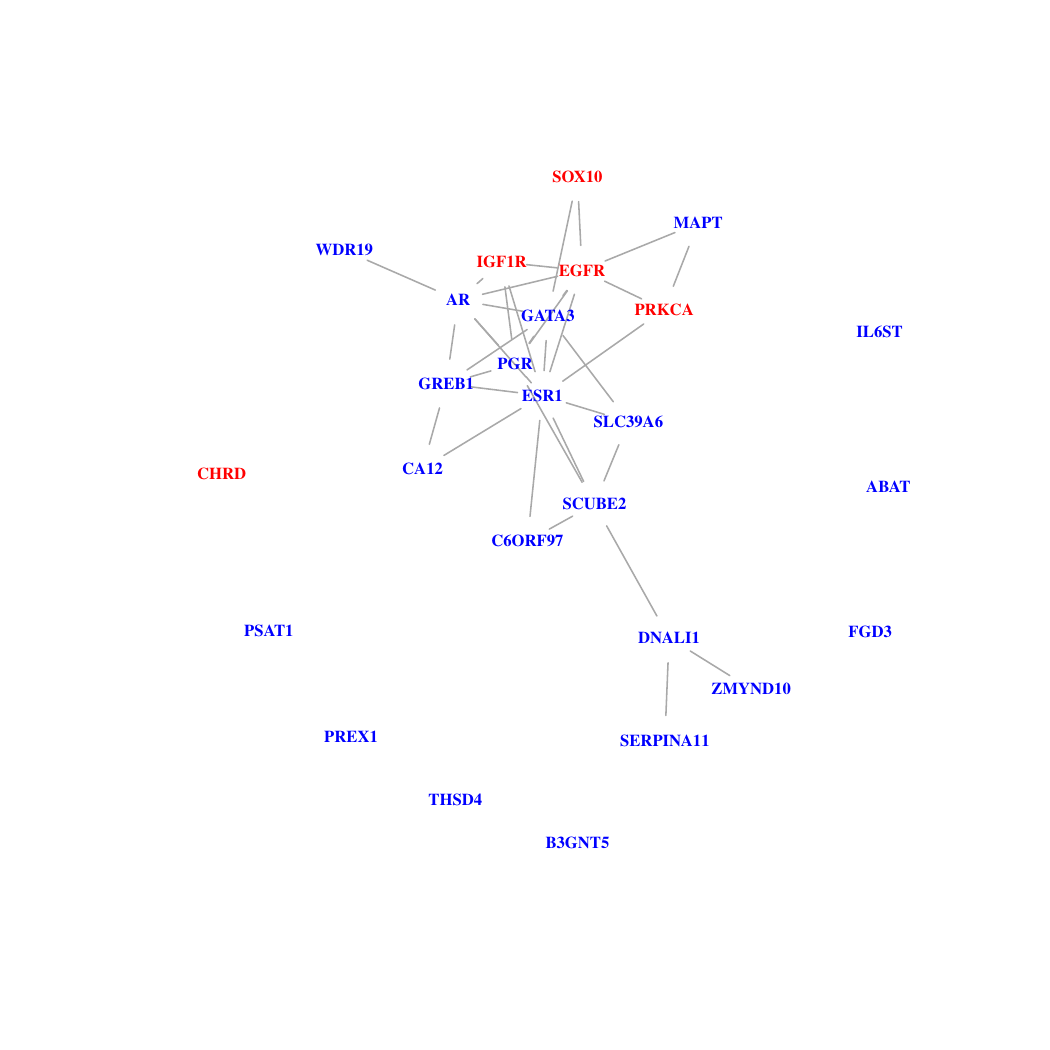}
	}
\vspace{-0.3in}
	\subfloat[]{%
		\includegraphics[width=0.85\columnwidth]{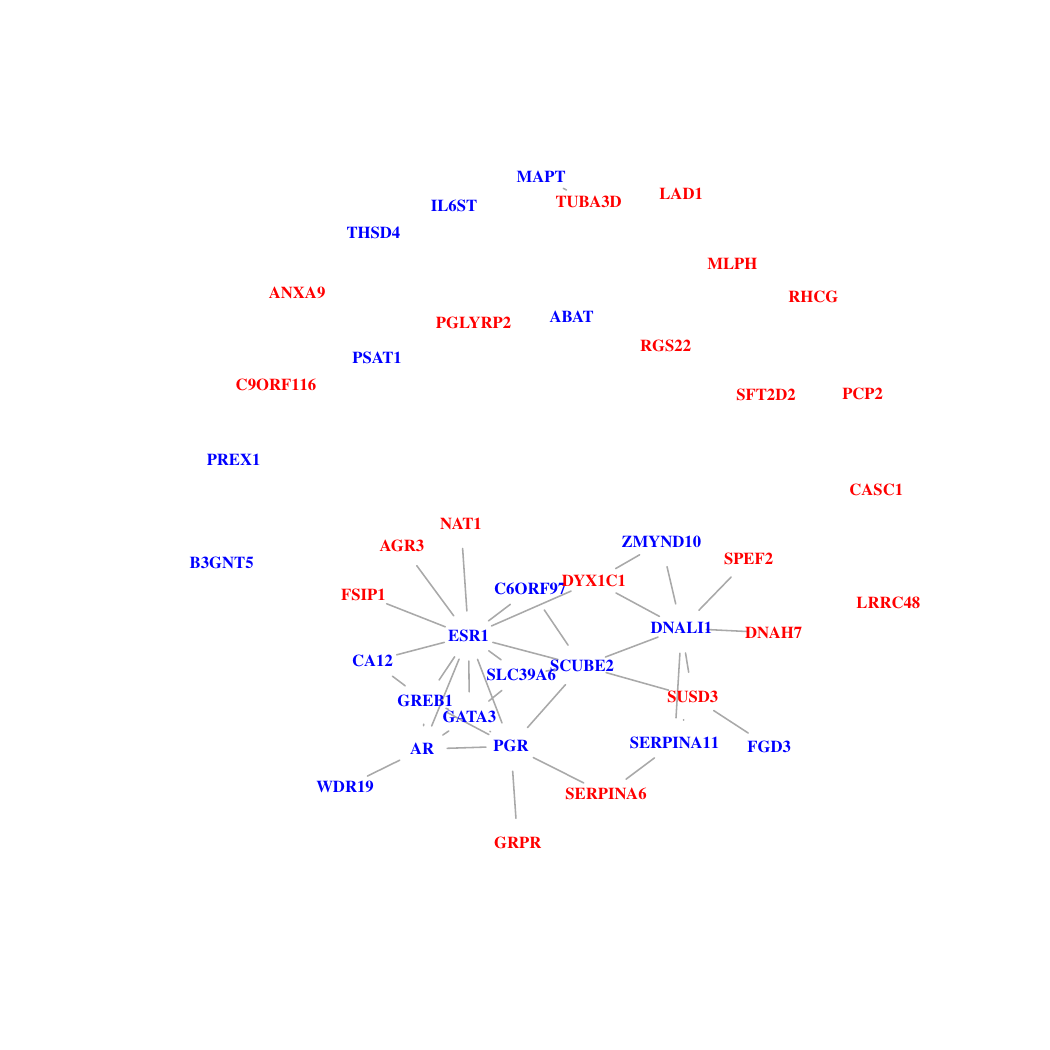}
	}
	
	\caption{(a) Induced networks of common top genes selected by network-guided RFs on both datasets. (b) Induced networks of common top genes selected by methods not using network information on both datasets. The genes in blue are those shown in Figure \ref{fig:exp_network_common_all} and genes in red are additional genes selected by each approach.}
	\label{fig:exp_network_w_wo_network}
\end{figure}

Next we investigate the common top genes selected by network-guided RFs and by methods not using network information separately to see the differences that network information may bring. The selected genes (including the aformentioned 21 genes) and their induced networks are shown in Figure \ref{fig:exp_network_w_wo_network}. We highlight these identified genes by two different colors. Genes in blue are those 21 genes shown in Figure \ref{fig:exp_network_common_all}. Genes in red are those identified only by each approach on both datasets.

When we look at the network of top genes identified by network-guided RFs, we observe that genes Epidermal Growth Factor Receptor (\emph{EGFR}) and Insulin Like Growth Factor 1 Receptor (\emph{IGF1R}) are added to the core module shown in Figure \ref{fig:exp_network_common_all}. Gene \emph{EGFR} belongs to Gene expression (Transcription) pathway and gene \emph{IGF1R} is in the ESR-mediated signaling pathway. Furthermore, the inclusion of \emph{EGFR} also builds a link between the core module to the Microtubule Associated Protein Tau (\emph{MAPT}) gene which is among the top important genes selected by all methods on both datasets. In Figure \ref{fig:exp_network_common_all}, gene \emph{MAPT} is an isolated node, but network-guided RFs include \emph{EGFR} into the set of important genes which further links to the gene \emph{MAPT} and strengthens the module.

In contrast, the methods without using network information do not identify any other genes from two PGR-related pathways. They select Gastrin Releasing Peptide Receptor (\emph{GRPR}) gene which has recently been found to be associated with the estrogen receptor (ER) positivity \cite{morgat2017gprp} and gene Serpin Family A Member 6 (\emph{SERPINA6}) which is identified as a marker of resistance to neoadjuvant chemotherapy in HER2-negative breast cancer \cite{de2013serpina6} (HER2 is short for Human epidermal growth factor receptor 2). More interestingly, the inclusion of gene Sushi Domain Containing 3 (\emph{SUSD3}) builds links between gene FYVE, RhoGEF And PH Domain Containing 3 (\emph{FGD3}) and the core module. Gene \emph{SUSD3} was shown to be highly expressed in ER$\alpha$-positive breast cancer tumors \cite{moy2015susd3}. On the contrary, even though the inclusion of gene Dynein Axonemal Assembly Factor 4 (\emph{DYX1C1}, also referred as \emph{DNAAF4}) does creates links between existing genes to result in a more connected module, this gene is more well-known for its association with deficits in reading and spelling ability \cite{bates2010dyslexia}. Apart from these genes, many isolated genes are included and the strengthening of the core module is very limited.

From this comparison, we can see that network information can be used in disease gene discovery and the information is likely to lead to a small amount of additional associated genes that can strengthen the core disease module. While if network information is not used, more isolated genes may be selected which could lead to challenges in understanding their contributions to the disease phenotype.

\section{Discussion}

In this paper, we systematically evaluate the performance of network-guided RF in terms of variable selection and prediction. The network-guided RF incorporates external network information of predictors into RF construction which is believed to be beneficial to disease gene and module identification. Our results suggest that for disease prediction, network information does not add much value. In terms of disease gene discovery, when disease genes are randomly distributed within the network, network information only deteriorates the gene selection, but if they form disease module(s), network-guided RF can identify disease genes and module(s) more accurately. We also find that when disease status is independent from genes in the given network, spurious gene selection results can occur when using network information, especially on hub genes. This phenomenon needs serious attention because hub genes often connect to many pathways. False selection of these genes may lead to intuitive but false conclusions. In the experimental data analysis, our results indicate that network-guided RF may lead to relevant gene identification which can further strengthen the core disease module. This demonstrates the potential gain on disease gene and module discovery.

However, a major limitation of the network-guided RF approach is that the gene selection procedure depends on a manually selected threshold for defining the top most important genes. It would be important to develop an automated variable selection procedure for RFs where the sampling probability of predictor variables is not uniform. Unlike in the commonly adopted procedures such as Boruta \cite{kursa2010boruta} or Vita \cite{janitza2018vita}, with a non-uniform sampling probability, the  (approximate) null distribution of variable importance measures is currently unknown and needs further study in order to automate the variable selection procedures. We would also like to point out that it is even harder to consider shadow variables as in the AIR \cite{nembrini2018air}, because introducing more predictor variables will inevitably change the sampling probability, thus may affect not only the null distribution construction but also the estimation of the importance measure itself.

In the construction of RF, the sampling probability of predictors can be considered as prior knowledge on the importance of predictors. From Bayesian perspective, the sampling probability used in the standard RF is a non-informative prior while the one based on network information is informative. Our simulation studies, if viewed from a Bayesian perspective, demonstrates the result of potential prior-data conflict, especially in the scenario where disease genes are randomly distributed in the network. Even though the practical relavence of this particular scenario may be limited, from methodology development perspective, it would be interesting to consider a more robust prior such that when there is any disease module, the network information can provide enough information to benefit the detection and when there is no disease module, such information should not hinder the data-driven detection, especially not leading to consistent false selection.

Our approach by using microarry and RNA-Seq datasets for validation in the experimental analysis may not be ideal. As pointed out in \cite{wang2014concordance} and \cite{mantione2014comparing}, the concordance between datasets coming from these two technologies may not be perfect and could be affected by a range of chemical treatment conditions. Therefore, it would be beneficial to apply network-guided RF on a large RNA-Seq dataset with carefully constructed networks.

We are aware that other approaches to incorporate network information have been published. Guan et al. (2018) \cite{guan2020knowledge} provide a general framework for incorporating prior knowledge into RF construction for biomarker discovery where network information may be considered. Zhao et al. (2020) \cite{zhao2020netembed} use network information in the feature engineering step and construct standard RF with these created features. Similarly, Adnan et al. (2019) \cite{adnan2020edge} propose to use network edges as features in RF construction to obtain a better predictive model. Thus, it would be interesting to conduct a comparative study to compare different approaches in terms of disease gene discovery. We will leave this work for a future project.

\section*{Acknowledgements}

This work was supported by the German Federal Ministry of Education and Research (BMBF) funded e:Med Programme on systems medicine [grant 01ZX1510 (ComorbSysMed) to SSzy]. The results published here are in part based on data generated by the TCGA Research Network: http://cancergenome.nih.gov/.

\section*{Conflict of interests}

The authors declare no conflict of interests.

\section*{Data availability}

An R package \texttt{networkRF} implementing the network-guided RF approach and providing functions for simulation is available at GitHub (https://github.com/imbs-hl/networkRF). The processed and quality controlled experimental datasets are also included in the R package.

\bibliographystyle{unsrt}
\bibliography{references}

\end{document}